\documentclass[useAMS,usenatbib]{mn2e}
\usepackage{graphicx}

\title[Emergent Universe from A  Matter-Energy Composition]{Emergent Universe from A  Composition of Matter, Exotic Matter and Dark Energy
}
\author[B. C. Paul, S. Ghose and P. Thakur]
  {B. C. Paul,$^{1,}$$^3$ \thanks{ Electronic mail:  bcpaul@iucaa.ernet.in}
   S. Ghose,$^{1,}$$^3$ \thanks{ Electronic mail:souvik@bose.res.in }
   P. Thakur$^{2,}$ $^3$ \thanks{ Electronic mail :  prasenjit \textunderscore thakur1 @yahoo.co.in} 
   \\
  $^1$Physics Department, North Bengal University \\ 
      Dist. : Darjeeling, Pin : 734 013, West Bengal, India 
\\
  $^2$Physics Department, Alipurduar College\\
      Dist. : Jalpaiguri, Pin : 736122, West Bengal, India
 \\
 $^3$ IUCAA Reference Centre, Physics Department \\
 North Bengal University \\}
\date{}
\begin{document}


\pagerange{\pageref{firstpage}--\pageref{lastpage}} \pubyear{2008}

\maketitle

\label{firstpage}
\begin{abstract}

A specific class of flat Emergent Universe (EU) is considered and its viability is tested in view of the recent observations. Model parameters are constrained from Stern data for Hubble Parameter and Redshift ($H(z)$ vs. $z$) and from a model independent measurement of BAO peak parameter.  It is noted that a composition of Exotic matter, dust and dark energy, capable of producing an EU, can not be ruled out with present data. Evolution of other relevant cosmological parameters, viz. density parameter ($\Omega$), effective equation of state (EOS) parameter ($\omega_{eff}$) are also shown.  

\end{abstract}


\section{Introduction}

It is known from observational cosmology  that our universe is passing through a phase of acceleration. Unfortunately, the  present phase of acceleration of the universe is not clearly understood. Standard Big Bang cosmology with perfect fluid assumption fails to accommodate the observational fact. However, an accelerating universe is permitted if a small cosmological constant ($\Lambda$) be included in the Einstein's gravity . There is, however, no satisfactory theory that explains the origin of $\Lambda$ which is required to be unusually small.  Moreover, Standard Big Bang model without a cosmological constant is inevitably pleagued with a time like singularity in the past. 
The Big Bang model is also found to be entangled with some of the observational features  which do not have explantion in the framework of perfect fluid model. Consequently an inflationary epoch in the early universe is required \citep{alt4} to resolve the outstandinng issues in cosmology.
It is not yet understood when and how the universe entered the phase. However, the concept of inflation is taken up to build a consistent scenario of the ealy universe. Inflation may be realized in a semiclassical theory of gravity where one requires an additional inputs like existence of a scalar field which describes the matter in the universe. An alternative approach is also followed where gravitational sector of the Einstein field equation is modified by including higher order terms in the Einstein-Hilbert action \citep{b19}. To address the present accelerating phase of the universe once again attempts are made where theories with a modification of the gravitational sector taking into account higher order terms that are relevant at the present energy scale are considered. There are other approaches generally adopted considering modification of the matter sector by including  very different kind of matter known as exotic matter namely, Chaplygin gas and its variations \citep{b3,b4}, models consisting one or more scalar field and tachyon fields \citep{b18}.  
While  most of these models address dark energy part of the universe,   other models based on non-equilibrium thermodynamics and Boltzmann formulation, which do not require any dark energy \citep{b15,b16,b17}, are also considered suitable for describing late universe. A viable cosmological model should accommodate an inflationary phase in the early universe with a suitable accelerating phase at late time.
An interesting area of cosmology is to consider models which are free from the initial singularity also. Emergent Universe (EU) scenario is one of the well known choices in this field. EU models are proposed in different framework like Brans-Dicke theory \citep{b151}, brane world cosmology \citep{b1,b2,b5}, Gauss-Bonnet modified gravity \citep{b13}, loop quantum cosmology \citep{alt3} and standard General Relativity (GR) \citep{b12}. Some of these models are implemented in a closed universe \citep{alt2} while others in a flat universe \citep{b12}. If EU be developed in a  consistent way it might solve some of the well known conceptual problems  not understood in the Big-Bang model. An interesting class of EU model in the standard GR framework has been obtained by \citet{b12} considering a non-linear equation of state in a flat universe.  The EU model evolves from a static phase in the infinite past into an inflationary phase and finally it admits an accelerating phase at late time. The universe is free from  initial singularity and  large enough to begin with so as to avoid quantum gravity effects. The non-linear equation of state is the input of the model which permits different composition of matter  in addition to normal matter as cosmic fluid. The model has been explored in a flat universe as such universe is supported by recent observations.  The EOS considered in obtaining EU model by \citet{b12} is
\begin{equation}
\label{eos1}
p=A\rho-B\rho^\frac{1}{2},
\end{equation}
where $A$ and $B$ are unknown parameters with  $B>0$ always.  Different values of $A$ and $B$ corresponds to different composition of matters in the EU model. In the  literature \citep{w2},  similar kind of non-linear EOS has been considered as a double component dark energy model and analyzed to obtain acceptable values of model parameters. 
The EOS given by eq. (\ref{eos1}) is a special form of a more general EOS, $p=A\rho -B\rho^{\alpha}$; which permits Chaplygin gas as a special case (with $\alpha <0$) \citep{b3,b4}. Chaplygin gas is considered widely in recent times to build a consistent cosmological model. It  interpolates between a matter dominated phase and a de Sitter phase. Later various modified forms of  Chaplygin gas were proposed \citep{b10} to track cosmological evolution. For example models like Modified Chaplygin gas interpolates between radiative era and $\Lambda$CDM era. \citet{w2} showed in their work that  such interpolation is permissible even with $\alpha>0$ and a string specific configuration may be phenomenologically realized with an EOS  considered by \citet{b12}. Recently using eq. (1) for an EU model proposed by \citet{b12},  we determined various constraints that are imposed on the EOS parameters  from observational data namely, SNIa data, BAO peak parameter measurement and CMB shift parameter measurement \citep{alt1}. It was noted that an EU  model is permitted with $A<0$. It is found that the possibility of $A=0$ case is  also permitted when we probe the contour diagram of $A-B$ plane with 95 $\%$ confidence. The case $A=0$ corresponds to a composition of dust, exotic matter and dark energy  in  the universe which is certainly worth exploring.  
In this paper a specific EU model is taken up where the matter energy content of the universe comprises of dust, exotic matter and dark energy. Using Stern data (Table. 1), the admissibility of model parameters are determined from $H(z)$ vs. $z$ \citep{b14} and using measurement of model independent BAO peak parameter $\mathcal A$. We also plotted evolution of cosmologically relevant parameters in our model. The paper is organized as follows : in section 2  field equations for the model are discussed, in section 3 and 4 we the model is constrained with Stern data and Stern+BAO data respectively. Finally in section 5 the findings are summarized with a discussion.

 \begin{table}
  \begin{minipage}{140mm}
  \caption{Stern Data ($H(z) vs. z$)}
  \begin{tabular}{l|c|r}
  \hline
  {\it z Data} & $H(z)$ & $\sigma$ \\
  \hline
   0.00 & 73  & $ \pm $ 8.0	 \\
   0.10 & 69  & $ \pm $ 12.0 \\
   0.17 & 83  & $ \pm $ 8.0 \\
   0.27 & 77  & $ \pm $ 14.0 \\
   0.40 & 95  & $ \pm $ 17.4 \\
   0.48 & 90  & $ \pm $ 60.0 \\
   0.88 & 97  & $ \pm $ 40.4 \\
   0.90 & 117 & $ \pm $ 23.0 \\
   1.30 & 168 & $ \pm $ 17.4 \\
   1.43 & 177 & $ \pm $ 18.2 \\
   1.53 & 140 & $ \pm $ 14.0 \\
   1.75 & 202 & $ \pm $ 40.4 \\
   
\hline
\end{tabular}
\end{minipage}
\end{table} 

\section{ Field Equations}
We consider Robertson-Walker(RW) metric which is given by :
\begin{equation}
\label{frw}
ds^{2} = - dt^{2} + a^{2}(t) \left[ \frac{dr^{2}}{1- k r^2} + r^2 ( d\theta^{2} + sin^{2} \theta \;
d  \phi^{2} ) \right]
\end{equation}
where  $k=0,+1(-1)$ is the curvature parameter in the spatial section representing flat or closed (open) universe and $a(t)$ is the scale factor of the universe, $r,\theta,\phi$ are the dimensionless comoving co-ordinates. The Einstein field equation is
\begin{equation}
\label{ef}
R_{\mu \nu} -\frac{1}{2} g_{\mu \nu} R = 8 \pi G T_{\mu \nu}
\end{equation}
where $R_{\mu \nu}$, $R$ and $T_{\mu \nu}$ represent Ricci tensor, Ricci scalar and energy momentum tensor respectively. Using RW metric  in Einstein  field equation we obtain time-time component which is ghiven by
\begin{equation}
\label{fr1}
3 \left( \frac{\dot{a}}{a} + \frac{k}{a^2} \right) = \rho
\end{equation}
where we consider natural units i.e., $c=1$, $8 \pi G = 1$.
Another field equation is the  energy conservation equation which is given by 
\begin{equation}
\label{consv}
\frac{d\rho}{dt} + 3 H (\rho + p) = 0 ,
\end{equation}
where $p$, $\rho$ and $H$ are respectively pressure, energy density, Hubble parameter $\left( H = \frac{\dot{a}}{a} \right)$
The Hubble parameter ($H$) can be expressed in terms of redshift parameter ($z$) which is given by
\begin{equation}
\label{hz}
H(z)=-\frac{1}{1+z}\frac{dz}{dt}.
\end{equation}
Since the components of matter and dark energy (exotic matter) are conserved separately, we may use energy conservation equation together with EOS given by eq. (\ref{eos1}) to determine the energy density which is obtained on integrating eq. (\ref{consv}) :

\begin{equation}
\label{rhoa}
\rho_{emu}= \left[\frac{B}{1+A}+\frac{1}{A+1}\frac{K}{a^{\frac{3(A+1)}{2}}}\right]^{2},
\end{equation}
where  $K$ is an integration constant and for a consistent formulation of EU it is required to be positive definite \citep{b12}. It is evident that the energy density  $\rho$ contains three terms corresponding to three different  composition of fluids :
\begin{eqnarray}
\label{rho123}
\rho(z)= \left(\frac{B}{A+1}\right)^2+ \frac{2BK}{(A+1)^2}(1+z)^{\frac{3}{2}(A+1)}+ \nonumber \\
 \left(\frac{K}{A+1}\right)^2(1+z)^{3(A+1)}
\end{eqnarray}

In the above  the first term is a constant which  may be considered to describe energy density corresponding to dark energy. In a simpler form eq. (\ref{rho123}) can be written as:
\begin{equation}
\rho(z)=\rho_{const} + \rho_1 (1+z)^{\frac{3}{2}(A+1)}+\rho_2 (1+z)^{3(A+1)}
\end{equation}
we denote $\rho_{const} = \left(\frac{B}{A+1}\right)^2$, $\rho_1=\frac{2BK}{(A+1)^2}$ and $\rho_2=\left(\frac{K}{A+1}\right)^2$ denote energy densities for different fluid components at the present epoch among which $\rho_{const}$ denotes the constant component. We note that present value of densities depend on both $B$ and $K$. The Einstein field equation given by eq.(\ref{fr1})  can be rewritten  for a flat  universe ($k = 0$) as :

\begin{equation}
\label{frz}
H(z)^2 = H_{0}^2 \left( \Omega_{const}+\Omega_{1}(1+z)^{\frac{3}{2(A+1)}}  +\Omega_2 (1+z)^{3(A+1)}\right)
\end{equation}
where $\Omega=\frac{\rho}{\rho_c}$ s denote density parameters for corresponding fluid and $\rho_c=\frac{3 H_o^2}{8 \pi G}$ here is the critical density. 

\begin{figure}
\label{stern}
\includegraphics[width=240pt,height=200pt]{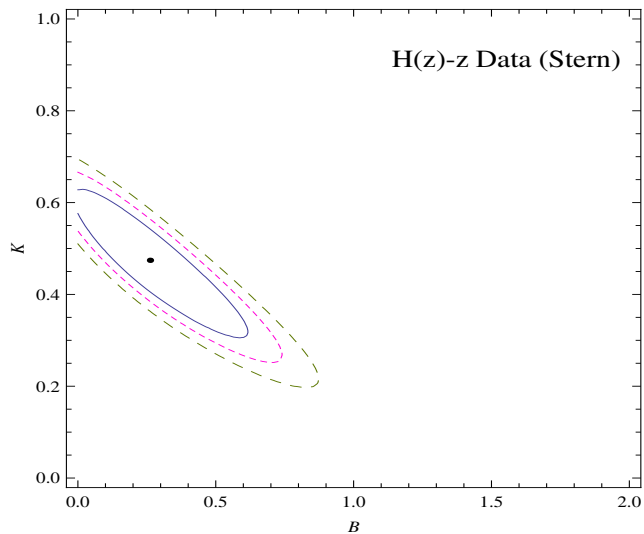}
\caption[Constraints from Stern Data]{(Colour Online)Constraints from Stern Data ($H(z) \: vs. \: z$)  $68.3\%$(Solid) $95\%$ (Dotted) and $99.7 \%$  (Dashing) contours. The best fit point is shown ($0.0122, -0.0823$).}
\end{figure}

\section{Constraints on model parameters from observational data}

In this section we consider an EU model  implemented in a flat universe using EOS given by eq.(\ref{eos1}). A special case  $A=0$ is taken up here to explore EU scenario with a definite composition of matter namely, dust, exotic matter and dark energy. In this case  eq. (\ref{frz}) can be represented in functional form given by 

\begin{equation}
\label{frz2}
H^{2}(H_{0},B,K,z )= H^{2}_{0}E^{2}(B,K,z),
\end{equation}
where
\begin{equation}
\label{fre}
E(B,K,z)^2 = \left( \Omega_{const}+\Omega_{1}(1+z)^{\frac{3}{2}}  +\Omega_2 (1+z)^{3}\right).
\end{equation} 
In the  present case $\Omega_{const}$ denotes a constant density parameter which corresponds to energy density described by a cosmological constant $\Lambda$. We denote the above density parameter by   $\Omega_{\Lambda}$. $\Omega_1$ corresponds to  some exotic matter which we denote by  $\Omega_e$. $\Omega_2$ corresponds to a dust like fluid which we  denote  by $\Omega_d$. Using the above  in  eq. (\ref{fre}) we obtain :
\begin{equation}
\label{fre2}
E(B,K,z)^2 = \left( \Omega_{\Lambda}+\Omega_{e}(1+z)^{\frac{3}{2}}  +\Omega_{d} (1+z)^{3}\right).
\end{equation}
The above functions will be used in the next section for analysis with observational results and to determine the model parameters.

\subsection{Analysis with Stern ($H(z) vs. z$) data}

\begin{figure}
\label{baof}
\includegraphics[width=240pt,height=200pt]{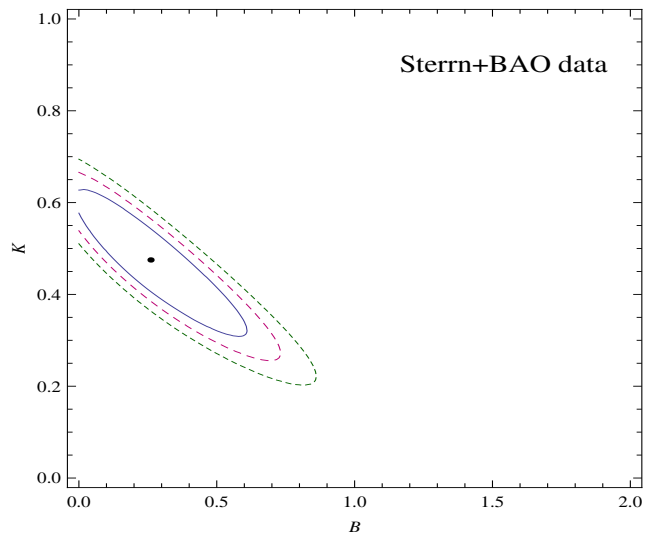}
\caption{(Colour Online)Constraints from joint analysis with Stern ($H(z)$-$z$) data and BAO peak parameter measurement for $68.3\%$(Solid) $95\%$ (Dashed) and $99.7 \%$ (Outermost) confidence level are shown in the figure along with the best fit value(0.0094,-0.1573)}
\end{figure}

In this section  we define $\chi^2$ function as :
\begin{equation}
\chi^{2}_{stern}(H_{0},A,B,K,z)=\sum\frac{(H(H_{0},B,K,z)-H_{obs}(z))^2}{\sigma^{2}_{z}}
\end{equation}
where $H_{obs}(z)$ is the observed Hubble parameter at redshift $z$ and $\sigma_{z}$ is the error associated with that particular observation. The present day Hubble parameter ($H_{0}$) is a nuisance parameter here. The objective of the analysis is to determine the constraints imposed on the model parameters namely, $B$ and $K$  from the observational input. So we can safely marginalize over $H_0$, defining a function\\
$
\nonumber L(A,B,K,z)= \int{Exp\left[-\frac{\chi ^{2}(H_0,A,B,K,z)}{2} \right]P(H_0) dH_0 }$ \\
where $P(H_0)$ represents a prior distribution function. Here we consider a Gaussian Prior with $H_0 = 72 \pm 8$. In the theoretical model it is demanded that the model parameters should satisfy the inequalities (i) $B>0$,(ii) $K>0$. Therefore, the model parameters obtained from the best fit analysis with observational data are determined in the theoretical parameter space. 
The best fit values obtained for the parameters here are: $B=0.2615$ and $K=0.4742$ together with  $\chi_{min}^2=1.02593$ ( per degree of freedom). The plots of $68.3 \%$, $95 \%$ and $99.7 \%$ confidence level contours are shown in fig. \ref{stern}. The following range of values are permitted : $0.003<B<0.5996$ and $0.303<K<0.63$ within $68.3 \%$ confidence level.

\subsection{Analysis with Stern+BAO data}

For a flat universe BAO peak parameter may be defined as in a low redshift region such as $0<z<0.35$ \citep{b6}:
\begin{equation}
\label{baod}
\mathcal {A} =\frac{\sqrt{\Omega_{m}}}{E(z_{1})^{1/3}}\left(\frac{\int ^{z_1}_0 \frac{dz}{E(z)}}{z_{1}}\right)^{2/3} 
\end{equation} where $\Omega_{m}$  is the total density parameter for matter content of universe. One has to consider a constant $\omega$ (EOS parameter). Even if $\omega$ is not strictly constant, it is quite reasonable to take a constant $\omega$ value over a small redshift interval. It would not be strictly the value of $\omega$ at $z=0$ but rather some average value in the region $0<z<0.35$. Here we use a technique adopted by \citet{b6} to explore the parameter $\mathcal {A}$ which is independent of dark energy model. The value of $\mathcal {A}$  for a flat universe is  $\mathcal {A} =0.469 \pm 0.017$ as measured in \citet{b6} using SDSS data. We define $\chi ^2_{BAO} = \frac{(\mathcal {A} - 0.469)^2}{(0.017)^2}$. For a joint analysis scheme we consider $\chi_{tot}^2= \chi_{stern}^2+\chi_{BAO}^2$. The best fit values found in the joint analysis are : $B=0.2599$ and $K=0.4751$ along with a $\chi_{min}^2=1.1681$ (per degree of freedom). Contours of $68.3 \%$, $95 \%$ and $99.7 \%$ confidence level are shown in fig. 2. Here we found that the range of values permitted within $68.3 \%$ confidence is a bit elevated : $0.009<B<0.606$ and $0.3126<K<0.6268$.

\section{Relevant cosmological parameters}

The range of permitted values of model paramters $B$ and $K$ are determined above. In this section we determine the variation of both the density parameter  and   the effective equation of state. Note that the model is an asymptotically de Sitter model and a late time phase of acceleration is assured. $68.3 \%$, $95 \%$ and $99.7 \%$ confidence level contours in $\Omega_d - \Omega_e$ plane are shown in fig. 3.
\begin{figure}
\label{omegaf}
\includegraphics[width=240pt,height=200pt]{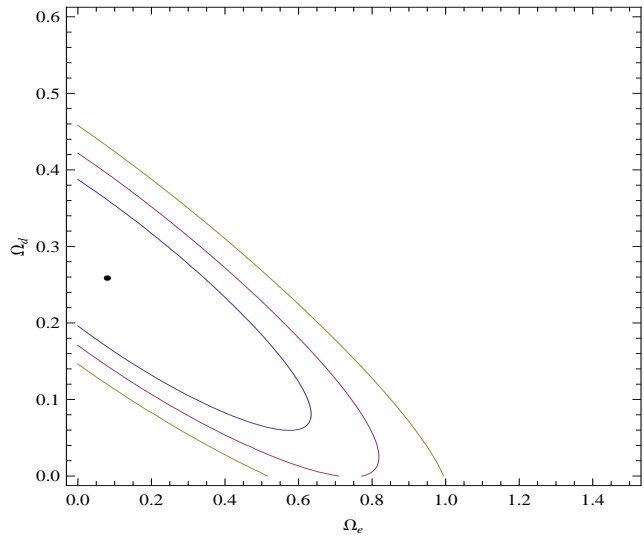}
\caption{(Colour Online)$68.3 \%$ (inner), $95 \%$ (middle) and $99.7 \%$ (outer) confidence level contours in $\Omega_d - \Omega_e$ plane.} 
\end{figure} 
It is found that the best fit value ($\Omega_d + \Omega_e =0.3374$) permits $\Omega_{\Lambda}=0.6626$. It is also noted that the generally predicted values $\Omega_{\Lambda}  \approx 0.72$ and $\Omega_d \approx 0.04$ are permitted here within $68.3 \%$ confidence level. We also plot the evolution of the effective EOS in fig. 4.   As expected it remains negative throughout. EU, as we have noted, is an asymptotically de Sitter universe and it is evident from fig. 5 that  $\rho$ decreases to a very small value at late universe.

\section{Discussions}

\begin{figure}
\label{effeos}
\includegraphics[width=240pt,height=200pt]{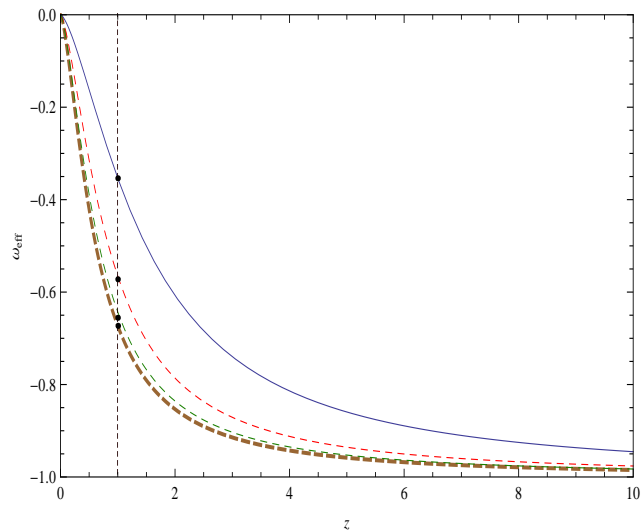}
\caption{(Colour Online) Evolution of $\omega_{eff}$ with best fit values and values within different confidence level}
\end{figure}

\begin{figure}
\label{rhop}
\includegraphics[width=240pt,height=200pt]{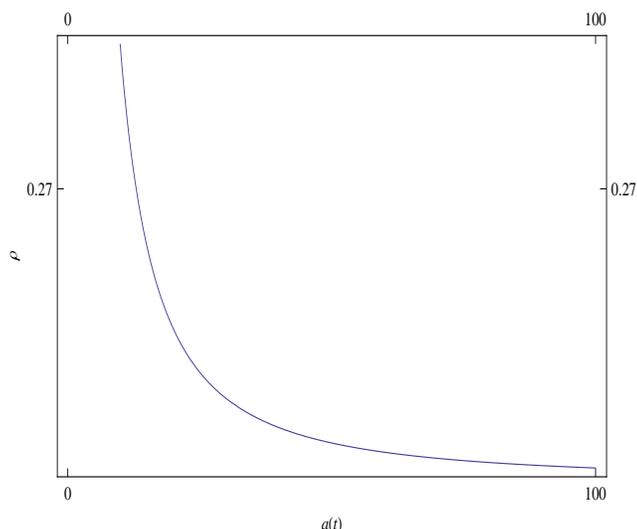}
\caption{(Colour Online) Evolution of the matter-energy density in a asymptotically de Sitter universe} 
\end{figure}

In this paper considering a very specific model of flat EU, we determine the observational constraints on the model parameters. For this recent observational data namely, Stern data, measurement of BAO peak parameter are used. The specific form of EOS given by eq. (1) to obtain EU scenario in a flat universe is employed here for the purpose. We set $A=0$ in the eq. (1) to begin with. $A$  equal to zero represents a universe with a composition of  exotic matter only. This kind of EOS has been considered in \citet{noz}. In our previous work \citep{alt1} on  EU model it is noted that a small non zero value of $A$ (although zero is not ruled out) is permitted. As a result the analysis was done with non zero $A$. In this paper since we are interested in a  specific composition of matter energy content of the universe corresponding to $A=0$ anlysis is carried out for EOS given by (1) with $A=0$ only. 
As suggested by \citep{b12},  it corresponds to the content of the universe which is a composition  of dark energy, dust and exotic fluid> The above composition is  reasonable to obtain a viable scenario of the universe considering the  observational facts. It seems that the exotic part of the EOS may also contribute in the budget of dark energy content of the universe. We found that the observationally favoured amount of dark energy present in universe today $\Omega_{\Lambda} \approx 0.72$ is permitted in our model within $68.3 \%$ confidence level.  However, it may be mentioned here that the model may be extended even if $ |A|<<1$ and so that we can write $A+1 \approx 1$. We found that a  composition of dust, exotic matter and dark energy may produce an EU model within the framework of Einstein's gravity with a non-linear equation of state.  It is also noted that this kind of model can accommodate many other composition of matter energy depending on the value of $A$. The viability for those will be taken up elsewhere.

\section*{Acknowledgments}
BCP and PT  would like to thank {\it IUCAA Reference Centre}, Physics Department, N.B.U for extending the facilities of research work. SG would like to thank CSIR for awarding Senior Research Fellowship. BCP would like to thank University Grants Commission, New Delhi for Minor Research Project (No. 36-365/2008 SR). SG would like to thank Dr. A. A. Sen for valuable discussions and  for hospitality at Centre for Theoretical Physics,  JMI, New Delhi.

\end{document}